\documentclass{iopjournal}
\usepackage{graphicx}
\usepackage{xcolor}
\usepackage{amsmath,amssymb}
\usepackage{float}
\usepackage{physics}
\usepackage{amsfonts}
\usepackage{verbatim}
\usepackage{balance}
\usepackage{endnotes}
\usepackage{footnote}
\usepackage{adjustbox}
\usepackage{gensymb}
\usepackage{subfigure}
\usepackage{float}
\usepackage{epigraph}
\usepackage{xcolor}
\usepackage[utf8]{inputenc}
\usepackage{setspace}
\usepackage{lipsum}
\usepackage{mwe}

\newcommand{\beq}{\begin{eqnarray}}
\newcommand{\eeq}{\end{eqnarray}}

\usepackage{amsmath}

\usepackage[normalem]{ulem}

\begin{document}

\articletype{Article type}

\title{Dynamics and transport in LiF using classical and machine learned interatomic potentials}

\author{H. L. Devereux$^1$\orcid{0000-0001-9449-7350}, M. Withington$^1$\orcid{0009-0004-5638-7440}, A. M. Elena$^2$\orcid{0000-0002-7013-6670}, C. Cockrell$^{3,*}$\orcid{0000-0002-8501-7287}, K. Trachenko$^1$\orcid{0000-0003-4832-7101}}

\affil{$^1$School of Physical and Chemical Sciences, Queen Mary University of London, Mile End Road, London, E1 4NS, UK}

\affil{$^2$Scientific Computing Department, Science and Technology Facilities Council, Daresbury Laboratory, Keckwick Lane, Daresbury, WA4 4AD, UK}

\affil{$^3$Nuclear Futures Institute, Bangor University, Bangor, LL57 1UT, UK}

\affil{$^*$ Corresponding author. \ \email{c.cockrell@bangor.ac.uk}}

\keywords{molten salts, viscosity, machine learned interatomic potentials, molecular dynamics, phonons}

\begin{abstract}
We perform molecular dynamics simulations of molten Lithium Fluoride, \textcolor{black}{comparing the classical Buckingham and Born-Huggins-Mayer potentials with machine-learnt interatomic potentials - the MACE-MP-0b3 (small) foundation model potential, and a fine-tuned version thereof}. We find that the MACE-MP-0b3, out-of-the-box, is able to accurately reproduce the experimental shear viscosity across the liquid state comparable with polarisable potentials despite not itself being polarisable. Previous viscosities computed using classical potentials are under-predicted, which has previously been attributed to a suppressed melting temperature. Prompted by this we compare the melting points of these potentials. \textcolor{black}{We find that the machine-learnt potentials address the significant underestimation of the melting point of the classical potentials, though the refined potential and the foundation model differ significantly between themselves. We probe the microscopic structure and dynamics of each model to further elucidate the considerable change in viscosity between classical and machine-learnt potentials. We find that, despite their lower melting temperature, classical potentials produce considerably stiffer, higher frequency, single-particle and collective \textit{charge-transport} dynamics on account of their fixed charges. However, in the long-wavelength limit, the many-body machine-learnt potentials feature more elastic collective motion that governs the diffusion of \textit{momentum} and is therefore responsible for the enhanced shear viscosity}.
\end{abstract}

\section{Introduction}
Molten salts have received long term and recent interest due to their use in solar cells, as electrolytes and solvents \cite{papageorgiou1996performance}, and potential utilisation as coolant fluids in nuclear reactors \cite{Rosenthal1970,FRANDSEN2020,Allen2007,Tang2015}. Water is the usual coolant in these nuclear applications but suffers from a high-reactivity \cite{US_nuclear2011}. In addition to the use of molten salts as a coolant in the reactor, there are proposals to use the salt as a carrier for the fuel, by dissolving the fuel into the salt. This would allow for adjustments to be made without unloading the core and reduce the amount of nuclear waste generated \cite{Merle2009,Locatelli2013}. Molten salts can host pyroprocessing reactions with very high activity nucleides to increase fuel efficiency and reduce waste generation \cite{Salanne2008,Locatelli2013,Uozumi2021,Lee2011,Mitachi2022}. Furthermore, molten salts have seen use as thermal storage media to complement renewable energy sources, with research interest growing in recent years \cite{Carabello2021,Bhatnagar2022,Yang2010, bauer2021molten}. 

A thorough understanding of the thermodynamic and transport properties of molten salts and their mixtures will underlie the upscaling of the above processes to industrial scales, however this understanding is hindered by a poor theoretical understanding of the liquid state \cite{myreview, mybook, granato}, and the inadequacy of classical atomistic models in generating faithful thermophysical data \cite{proctor1, chen-review}. Determining thermophysical properties, such as shear viscosity, of molten salts accurately is crucial for designing the systems that use them \cite{nunes2003importance}. 

Shear viscosity is a key property governing the performance of working fluids in thermal hydraulics, which has been investigated theoretically in relation to its temperature dependent minimum that may be related to fundamental physical constants \cite{sciadv1, myreview}. We previously investigated the viscosity minimum in LiF using classical molecular dynamics (MD) \cite{Withington2024Viscosity} where, as found elsewhere \cite{LUO2016203}, the Buckingham potential model of molten LiF \cite{Buckingham1938, TOSI196445, Sangster1976Interionic, LUO2016203} exhibits a noticeable temperature offset compared to experimental results. Ciccotti \textit{et al.} reported a slightly better agreement using the Born-Huggins-Mayer (BHM) potential and by analysing the response to an explicit shearing force for one temperature point \cite{Ciccotti1976Transport} as far back as 1976. Despite these issues, due to their speed, the Buckingham and BHM potentials are still used for modelling LiF and other ionic systems \cite{galvez2018simulations, lv2018first, yang2024molecular, lu2021thermal, galashev2023molecular}. Some of these issues are addressed by polarisable potentials, such as the polarisable ion model (PIM), which are first-principles-parameterised using density functional theory (DFT) data specific to each application \cite{salanne2009heat, ishii2014thermal, gheribi2016formulation, gallagher2022investigation}. These are notably superior for determining thermophysical properties of ionic liquids e.g. LiF viscosity \cite{gheribi2014prediction}. More recently approaches using machine learning have also been developed in particular for LiF and FLiBe \cite{rodriguez2021thermodynamic, lam2021modeling}.

Motivated by these applications and a desire to improve the \textcolor{black}{understanding of molten LiF, both from theoretical and simulations points of view,} we now also turn to the rapidly growing field of machine-learnt interatomic potentials (MLIPs). The MACE-MP-0b3 (MACE) MLIP \cite{Batatia2022mace, Batatia2022Design,batatia2023foundation} is one example that is general purpose, transferable, and which has seen diverse success across atomistic material modelling, even without case-by-case parametrisation. \textcolor{black}{The MACE-MP-0b3 training set contains solid LiF, and a statement containing all the structures for LiF and other solids is available in Ref. \cite{Batatia2022mace}. paper in appendix B4} Here we report the accuracy of viscosity values calculated using the \textcolor{black}{MACE-MP-0b3 and a fine-tuned MACE potential\cite{tompa2026finetuningmlipfoundationmodels}} against experimental viscosity data\textcolor{black}{, and relate these values to the thermodynamics and dynamics of the liquid according to the phonon theory of liquid thermodynamics \cite{ropp,Cockrell2021b}.}

\section{Methods}
We perform MD simulations of molten LiF using classical MD (using the Buckingham and BHM potentials), \textit{via} the DL\_POLY package \cite{devereux2025dl_poly}, and MACE, \textit{via} the python package janus-core \cite{elliott_2024_14001356} using MACE-MP-0b3 (small flavour). We also report a fine tuned version \textcolor{black}{the potential against newer generation of MACE\cite{batatia2025crosslearningelectronicstructure}}  using the data of Fan \textit{et al.} \cite{fan2025multistage}, which we compare with PIM viscosity results\cite{gheribi2014prediction}. 

For the Buckingham and BHM potentials we use the standard parameters of Tosi and Fumi for alkali halides \cite{TOSI196445, Sangster1976Interionic, Ciccotti1976Transport, LUO2016203}. In DL\_POLY the Buckingham and BHM potentials have the form
\begin{align}
    U_\mathrm{Buckingham}(r) &= A e^{-\frac{r}{\rho}}-\frac{C}{r^6}\label{buck},\\
    U_\mathrm{BHM}(r) &= Ae^{B(\sigma-r)}-\frac{C}{r^6}-\frac{D}{r^8}\label{bhm},
\end{align}
for parameters $A, B, \sigma, \rho, C$, and $D$ and inter-atomic distance $r$. The parameters for all of the interactions are given in Tab. \ref{tab:buck}. Li and F were assigned their formal charges of +1 and -1 respectively. When performing our simulations in DL\_POLY we use the Smooth-particle Mesh Ewald method for electrostatic interactions \cite{essmann1995smooth}. \textcolor{black}{ A real space cutoff of 10 \AA \ was used for all classical MD integration.}
\begin{table}
    \centering
    \begin{tabular}{|l|c|c|c|}
    \hline
        Buckingham & Li-Li & Li-F & F-F \\\hline
        $A$ [eV] & 98.92 & 228.99 & 420.48 \\
        $\rho$ [\AA] & 0.299 & 0.299 & 0.299 \\
        $C$ [\AA$^{6}$ eV] &  0.046 & 0.499 & 9.051 \\\hline
        BHM & & & \\\hline
        $A$ [eV] & 0.422 & 0.290 & 0.158 \\
        $B$ [\AA$^{-1}$] & 3.344 & 3.344 & 3.344 \\
        $\sigma$ [\AA] & 1.632 & 1.995 & 2.358 \\
        $C$ [\AA$^{6}$ eV] & 0.0456 & 0.499 & 9.050 \\ 
        $D$ [\AA$^{8}$ eV] & 0.019 & 0.374 & 10.610 \\\hline
    \end{tabular}
    \caption{Buckingham and BHM potential parameters (in DL\_POLY format) corresponding to Equations \ref{buck} and \ref{bhm} to three decimal places.}
    \label{tab:buck}
\end{table}

The MACE model relies upon a Message Passing Neural-Network (MPNN) architecture where the input space is the dynamic local environment around any given atom along with its static properties (e.g. chemical element). Messages are constructed with radial basis functions and spherical harmonics to encode known symmetries in physical applications. Importantly the MACE model utilises higher-order many-body ACE expansion \cite{bernstein2024gap} for message passing to achieve faster and more accurate convergence to ground-state DFT energies, forces, and stresses than other similar MLIP architectures. The model is trained upon the MPtrj data set introduced for CHGNet \cite{deng2023chgnet} and compiled from the Material Project dataset \cite{jain2013commentary}, which includes 89 elements (including Li and F) in $\sim 146000$ materials. Not all the MLIPs are suitable for molecular dynamics studies since not all possess conservative forces and stresses. MACE, however, is a conservative model \cite{matbench}.

\textcolor{black}{Fine tuning was carried out using multihead strategy with matpes-pbe-omat as foundational head and using a replay dataset of 100000 structures randomly selected from the foundational model training set.  Fine-tuning runs over 20 epochs with a 1:10 ratio for energy and force weights, a learning rate of 0.0001, a decay of 0.99999 and batch size of 2. The rest of the settings were MACE defaults as of version 0.3.15. The training set contained the SCAN data from \cite{fan2025multistage}, containing 11377 structures, plus 1264 in validation 1264 and 3166 in testing. Final root-mean-square errors in energy in meV/atom are 3.7 for the training set, 7.8 for the validation set and 2.8 for the test set, while root-mean-square errors in forces in meV/\AA are 7.7 for the training set, 7.9 for the validation set and 7.8 for the test set.}

\textcolor{black}{The training dataset from \cite{fan2025multistage} was performed with SCAN functionals, whereas the MACE-MP-0b3 is with PBE functionals. The contribution of the functional choice to the difference between the foundation model and the fine-tuned model is beyond the scope of this study.}

For each potential, we use the Green-Kubo method \cite{Zwanzig1965, tuckerman2023statistical, zhang2015reliable} to calculate viscosity from the integral
\begin{equation}
    \eta=\frac{V}{k_{\mathrm{B}}T}\int_0^\infty \dd t \ \langle \sigma_{xy}(t) \sigma_{xy}(0) \rangle, \label{eq:visc}
\end{equation}
where $\langle \cdot \rangle$ is the ensemble average, $T$ is the mean system temperature and $k_\mathrm{B}$ is Boltzmann's constant. Equation \ref{eq:visc} may be calculated using DL\_POLY's on-the-fly correlator. The stress tensor is defined as
\begin{equation}
    \sigma_{\alpha\beta} = \frac{1}{V}\sum_{i}m^{i}v^{i}_{\alpha}v^{i}_{\beta} - \frac{1}{V}\sum_{i, j\neq i}r_{\alpha}^{ij}F_{\beta}^{ij}, \label{eq:stress}
\end{equation}
where $\alpha$ and $\beta$ are Cartesian component indices, $m^{i}$, $v^{i}$, $r^{i}$ are particle $i$'s mass, velocity, and position, and $\mathbf{r}^{ij} = \mathbf{r}^{i} - \mathbf{r}^{j}$, $\mathbf{F}^{ij}$ is the force on $i$ due to $j$ and $V$ is the system volume. Finally, $x$ and $y$ are orthogonal Cartesian coordinates. We also compute the average viscosity utilising the other off-diagonal terms, $z-x$ and $y-z$, analogously to Equation \ref{eq:visc}. We also use the python package ASE \cite{ase-paper} to calculate the partial radial distribution functions (RDFs), $g(r)$. 

For accurate viscosity statistics a long simulation run-time is required as well as a sufficient number of independent samples \cite{zhang2015reliable, LUO2016203, Cockrell2021Universal, Withington2024Viscosity}. For all statistics collection simulations we use the NVE ensemble with a timestep of $1$ fs and a trajectory length of 1 ns, and average across $20$ (for MLIPs) or 80 (for the classical potential) initial configurations each independently initialised with the same density. For equilibration we heat the system in the NPT ensemble from a crystal lattice structure taking temperature steps from $10$ K to the desired temperature in steps of $50$ K. Each subsequent heating step lasts for 10 ps, except when determining the melting point, in proximity to which we simulate for 50 ps. We perform a final 10 ps equilibration at each temperature with 20 different initial velocity seeds in the NVE ensemble before production runs. \textcolor{black}{We achieve a relative error of around 10\% using 20 trajectories of 1 ns, and around 5\% for 80 trajectories.}

Performed on GPUs using the janus-core package \cite{janus}, systems of 800 are forward evolved at a rate of around 0.095 seconds per timestep. This results in around 26 hours to complete a 1 nanosecond trajectory of 800 atoms on a single GPU. Using GPU nodes with four devices, the viscosity calculations for each of the four temperatures along the isobar can be therefore be performed in just over three weeks. Larger systems of 6400 atoms used for current calculations (see below) were integrated 7 times slower.

The viscosity calculated from Eq. \ref{eq:visc} is compared to experimental values \cite{janz1977physical, ABE1981173, Ejima1987ViscosityOM, ana-maria}. For both MD methods we use the same maximum correlation time for calculating $\eta$, $5$ ps. Janus-core is able to make use of MACE GPU hardware acceleration for calculations, however the added computational complexity of the many-bodied MACE potential stills leads to a far longer simulation wall-time (typically at least two orders of magnitude) than the CPU based DL\_POLY calculations based on classical potentials. In both we use a system size of $N=800$ atoms, where we have previously found little impact on results from utilising large system sizes up to 100000 atoms for classical potentials in LiF \cite{Withington2024Viscosity} and Argon \cite{Cockrell2021Universal} systems.

\textcolor{black}{We calculate the collective motion and dispersion curves using current correlation functions \cite{Boon1991}. The total mass current (momentum density) $\mathbf{J}(t, \mathbf{k})$ is given by 
\begin{equation}
    \label{eqn:masscurrent}
    \mathbf{J}_{\mathrm{T}}(t,\mathbf{k}) = \sum_{i=1}^{N} m_i \mathbf{v}_i(t) \exp \left( i \mathbf{k} \cdot \mathbf{r}_i(t)\right),
\end{equation}
which can be split into partial components $\mathbf{J}_1$ and $\mathbf{J}_2$ (with index 1 corresponding to Li, and 2 to F), such that 
\begin{equation}
    \label{eqn:totalmomentum}
    \mathbf{J}_{\mathrm{T}}(t,\mathbf{k}) = \mathbf{J}_1(t,\mathbf{k}) + \mathbf{J}_2(t,\mathbf{k}).
\end{equation}
In the hydrodynamic limit, $k \rightarrow0$, conservation of momentum dictates that $\mathbf{J}_1 = -\mathbf{J}_2$ in the barycentric frame of reference. In a two component system, there are two independent momentum density variables. Though we could nominate either partial mass current as the second independent variable, we instead construct the mutual mass current $\mathbf{J}_{\mathrm{M}}(t,\mathbf{k})$:
\begin{equation}
    \label{eqn:mutualmasscurrent}
    \mathbf{J}_{\mathrm{M}}(t,\mathbf{k}) = c_2 \mathbf{J}_1(t,\mathbf{k}) - c_1 \mathbf{J}_2(t,\mathbf{k}),
\end{equation}
with $c_I = \rho_I/\rho$, the mass fraction. In the hydrodynamic limit, $\mathrm{J}_X \rightarrow \mathbf{J_1}$, however $\mathrm{J}_X$ has the advantage of being orthogonal to $\mathrm{J}_T$ at all wavelengths. That is:
\begin{equation}
   \label{eqn:orthogonal}
   \left\langle \mathbf{J}_{\mathrm{T}}(t,\mathbf{k}) \cdot \mathbf{J}_\mathrm{M}(t, -\mathbf{k}) \right\rangle = 0.
\end{equation}
The mutual diffusion coefficient is calculated using the autocorrelation function of the mutual mass current \cite{Zhou1996}. We also note that the charge current $\mathbf{j}_q (t, \mathbf{k}) = q_1/m_1 \ \mathbf{J}_1 (t, \mathbf{k}) + q_2/m_2 \ \mathbf{J}_1(t, \mathbf{k})$, with $q_i$ the charge of the i$^{\mathrm{th}}$ ion species, is related to the mutual mass current \textit{via}:
\begin{equation}
    \label{eqn:chargemass}
    \mathbf{J}_{\mathrm{M}} (t, \mathbf{k}) = \left( \frac{m_1 m_2}{m_2 q_1 - m_1 q_2} \right) \mathbf{j}_q (t, \mathbf{k}),
\end{equation}
under the proviso of total charge neutrality: $N_1 q_1 + N_2 q_2 = 0$, with $N_i$ the number of ions of species $i$ in the system. The mass concentration current is therefore proportional to the charge current, and our orthogonal basis set is $\left\{ \mathbf{J}_T(t, \mathbf{k}), \mathbf{J}_X(t, \mathbf{k}) \right\}$, or equivalently $\left\{ \mathbf{J}_T(t, \mathbf{k}), \mathbf{j}_q(t, \mathbf{k}) \right\}.$ Each of these functions has transverse and longitudinal parts with respect to wavevector $\mathbf{k}$.}

\textcolor{black}{The time correlation functions available to this basis set are:
\begin{equation}
    \label{eqn:CTT}
    C_{\mathrm{TT}}(t, k) = \left\langle \mathrm{J}_\mathrm{T}(t,\mathbf{k}) \cdot \mathrm{J}_\mathrm{T}(0, -\mathbf{k}) \right\rangle,
\end{equation}
\begin{equation}
    \label{eqn:CTM}
    C_{\mathrm{TM}}(t, k) = \left\langle \mathrm{J}_\mathrm{T}(t,\mathbf{k}) \cdot \mathrm{J}_\mathrm{M}(0, -\mathbf{k}) \right\rangle = C_{\mathrm{MT}}(t, k) ,
\end{equation}
\begin{equation}
    \label{eqn:CMM}
    C_{\mathrm{MM}}(t, k) = \left\langle \mathrm{J}_\mathrm{M}(t,\mathbf{k}) \cdot \mathrm{J}_\mathrm{M}(0, -\mathbf{k}) \right\rangle,
\end{equation}
with each function depending on the magnitude of the wavevector $k = \abs{\mathbf{k}}$ and its transverse or longitudinal polarisation (denoted by superscripts $C^{\mathrm{t}}$ and $C^{\mathrm{l}}$ respectively) due to the isotropy of the liquid. The total mass current $\mathbf{J}_{\mathrm{T}}$ measures the fluctuations in mass density, so its spectrum $C(\omega,k)$ is related to stress-density waves in the hydrodynamic limit (\textit{i.e.} sound) - we therefore call it the acoustic spectrum. Meanwhile, the spectrum of the mutual mass current autocorrelation function couples to fluctuations in charge -  we therefore call it the optical spectrum. In each model at each state point, we sample wavevectors oriented along he $x$, $y$, and $z$ directions, with magnitudes equal to $2 \pi n/L$, where $n$ is an integer from 1 to 10 and $L$ is the width of the cubic cell. In order to access smaller wavenumbers, current calculations are performed in systems of 6400 atoms. Each autocorrelation function is calculated from a 40 ps trajectory, sampled every 2 fs, and averaged over 80 different initial conditions.}

\section{Results and Discussion}
\begin{figure}
    \centering    \includegraphics[width=0.85\linewidth]{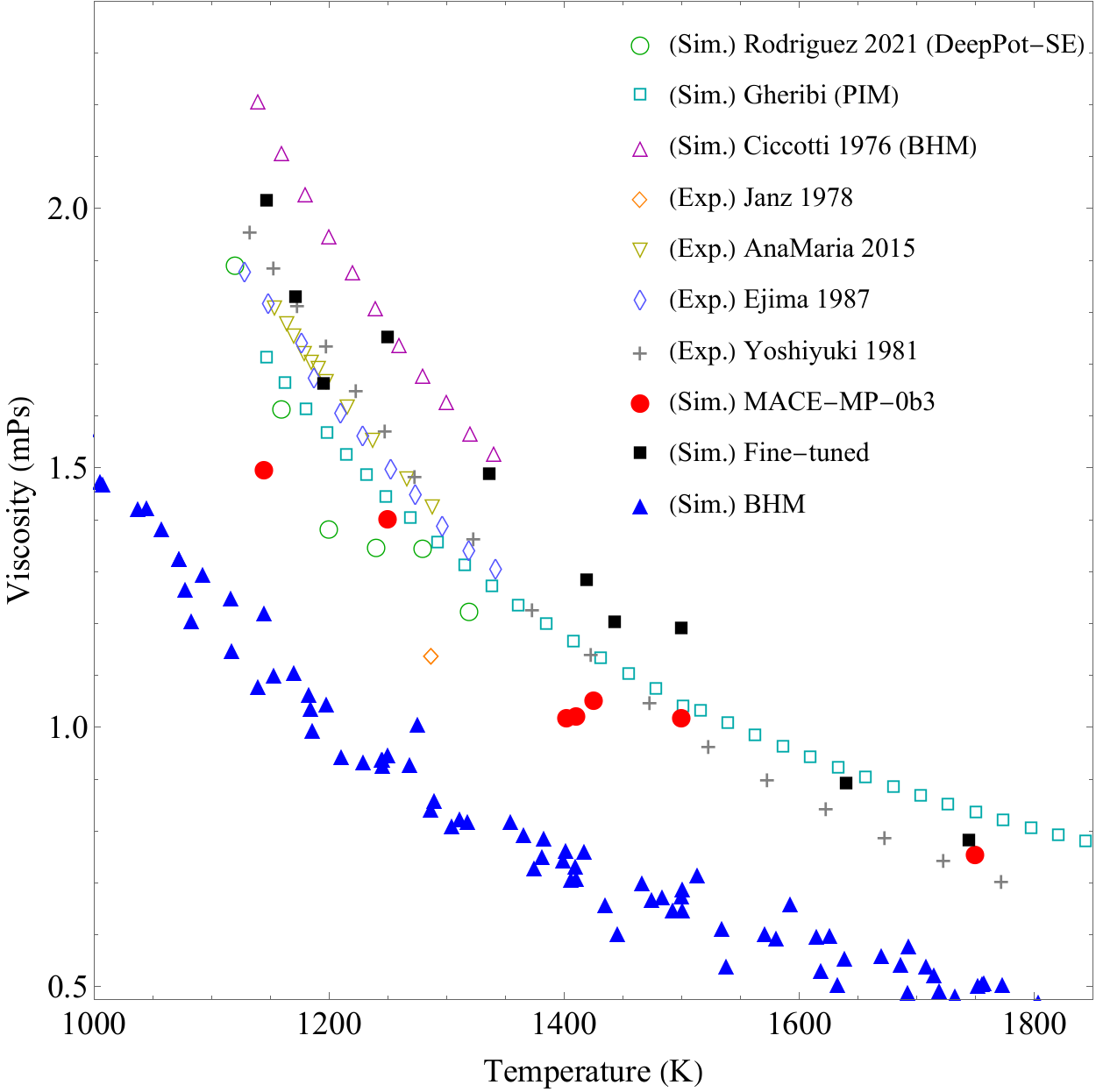}
    \caption{Viscosity, $\eta$. Plotted are experimental results \cite{janz1977physical, ABE1981173, Ejima1987ViscosityOM, ana-maria} and calculated results using MACE, BHM from Ciccotti \textit{et al.}, and our BHM results. We also show results using the polarizable ion model (PIM) from Gheribi \textit{et al.} \cite{gheribi2014prediction}. As reported previously for LiF \cite{LUO2016203, Withington2024Viscosity}, the classical potential has the correct trend but is off-set on the temperature axis.}
    \label{fig:viscosity}
\end{figure}

\begin{figure}
    \centering
    \includegraphics[width=0.8\linewidth]{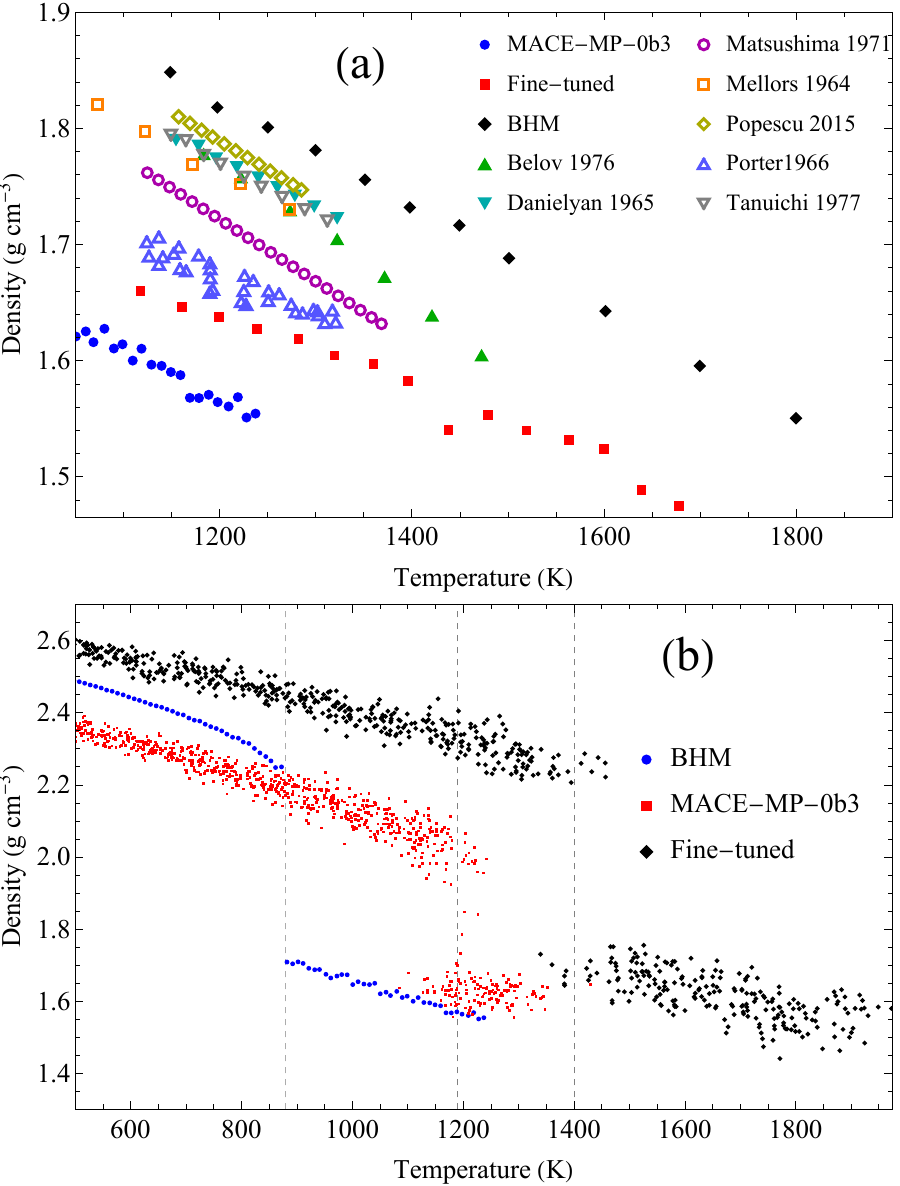}
    \caption{Volume from heating LiF from a crystal structure from $10$ K up to $1250$ K. The melting point is more accurately determined by the MACE potential than the Buckingham potential. CF the off setting of the shear viscosity data along the temperature axis in Figure \ref{fig:viscosity}. The offset of the melting point in temperature is approximately the same (200-300 K). The denser scatter points from 1000 K in the MACE\_MP\_0 results are an additional heating run with a finer temperature step.}
    \label{fig:melting}
\end{figure}

We begin by comparing the viscosity measurements from the three computational models to experimental data in Figure \ref{fig:viscosity}. \textcolor{black}{We recall the known offset} for the Buckingham and BHM models: the simulated viscosity matches the experimental viscosity well, albeit with the latter at a temperature around 300 K higher. \textcolor{black}{The MACE foundation model produces a viscosity on the lower end of the considerable spread in experimental data, a considerable improvement on the classical models.} The MACE and PIM results show similar accuracy, which is reasonable given that both are parametrised by DFT data. This does suggest that MACE could be used as an accurate general alternative for polarisable models of ionic systems even though it does not account for polarisation explicitly. \textcolor{black}{The fine-tuned model exhibits viscosity that meanwhile straddles top of the experimental range. This excession exhibited by the fine-tuned model may be partially explained by its increased density, as seen in Fig. \ref{fig:melting}a. Experimental densities around 1300 K are reported from around 1.6 - 1.85 g cm$^{-3}$ \cite{Birri2025} - the fine tuned model is therefore closest to these measurements, though there is considerable spread \cite{Birri2025}. A higher density will promote solid-like motion, increasing the local elasticity of the liquid structure and therefore better resist shear motion. While functional choice can have significant effect on density (and therefore viscosity) \cite{fan2025multistage}, a thorough investigation of this effect is beyond the scope of this study.}

Previously, the offsetting of viscosity values in the Buckingham results have been ascribed to the model undergoing melting at a suppressed temperature value commensurate with the gap between the simulated and experimental viscosity data \cite{LUO2016203}. This is a well known problem, with accurate melting temperatures generally determined using more complex methods such as solid-liquid phase co-existence including specifically for MD simulations of LiF \cite{belonoshko2000molecular}. Naively heating a crystal is, however, a common form of equilibration in general. To investigate this, in Fig. \ref{fig:melting} we show the system density for the BHM, MACE, and fine-tuned models following a gradual monotonic heating from a crystal structure. \textcolor{black}{We observe that the MLIPs do not exhibit the severely underestimated melting temperature of the classical models. The fine-tuned model's melting temperature exceeds that of the foundation model, the same trend seen in viscosity.} This gives some support to this notion that the poor reproduction of the experimental viscosity in simulations governed by the Buckingham and BHM models is related to the underestimation of the the melting temperature by approximately $300$ K, ascertained in experiments to be $1121$ K \cite{douglas1954lithium,lide2004crc} (although 1143 K is also recorded \cite{mead1974comparison}). \textcolor{black}{We furthermore observe a particularly} good agreement between the experimental melting temperature and the melting temperature of the MACE simulation, \textcolor{black}{while our fine tuned model overestimates by around 250 K. We note that, in the case of the fine-tuned MLIP, the lower temperatures in Figs. \ref{fig:viscosity} and \ref{fig:melting}a imply somewhat undercooling the liquid below its nominal melting point. As we note above, however, this may not represent a true thermodynamically supercooled liquid.} 


However, it should be noted that there are several factors contributing to the location of melting point in molecular simulations. For example, simulated systems often employ periodic boundary conditions and hence have no free surfaces or interfaces, present experimentally, which promote melting. Another factor is the small system sizes in MD simulations compared to experiments. This leads to the absence of long-wavelength and low-frequency phonons with large displacement amplitudes which destabilise the solid structure and likewise encourage melting. For these and other reasons, gradual heating from a crystal structure will not generally be accurate for determining the melting point \cite{corradini2014effect, lanning2004solid,frenkel2002understanding} and other methods such as calculating the free energies of solid and liquid phases, simulating phase co-existence and so on \cite{alfe2002complementary} can be used instead. \textcolor{black}{For this reason, we may infer that the MACE-MP-0b3 model likely still underestimates the temperature of melting, and our fine-tuned model may or may not improve on this underestimation. A more rigorous study of the thermodynamics and kinetics of liquid nucleation would be needed to address these inferences and is beyond the scope of this study. An existing study on the melting of LiF using ACE-scan MLIPs found a good agreement between simulated and experimental melting temperature \cite{fan2025multistage}. Overall, then, we acknowledge a significant improvement on the melting point estimation when using MLIPs instead of classical fixed-charge potentials, and note the correlation of this improvement with the improved viscosity predictions.}


\begin{figure*}
    \centering
    \includegraphics[width=0.9\linewidth]{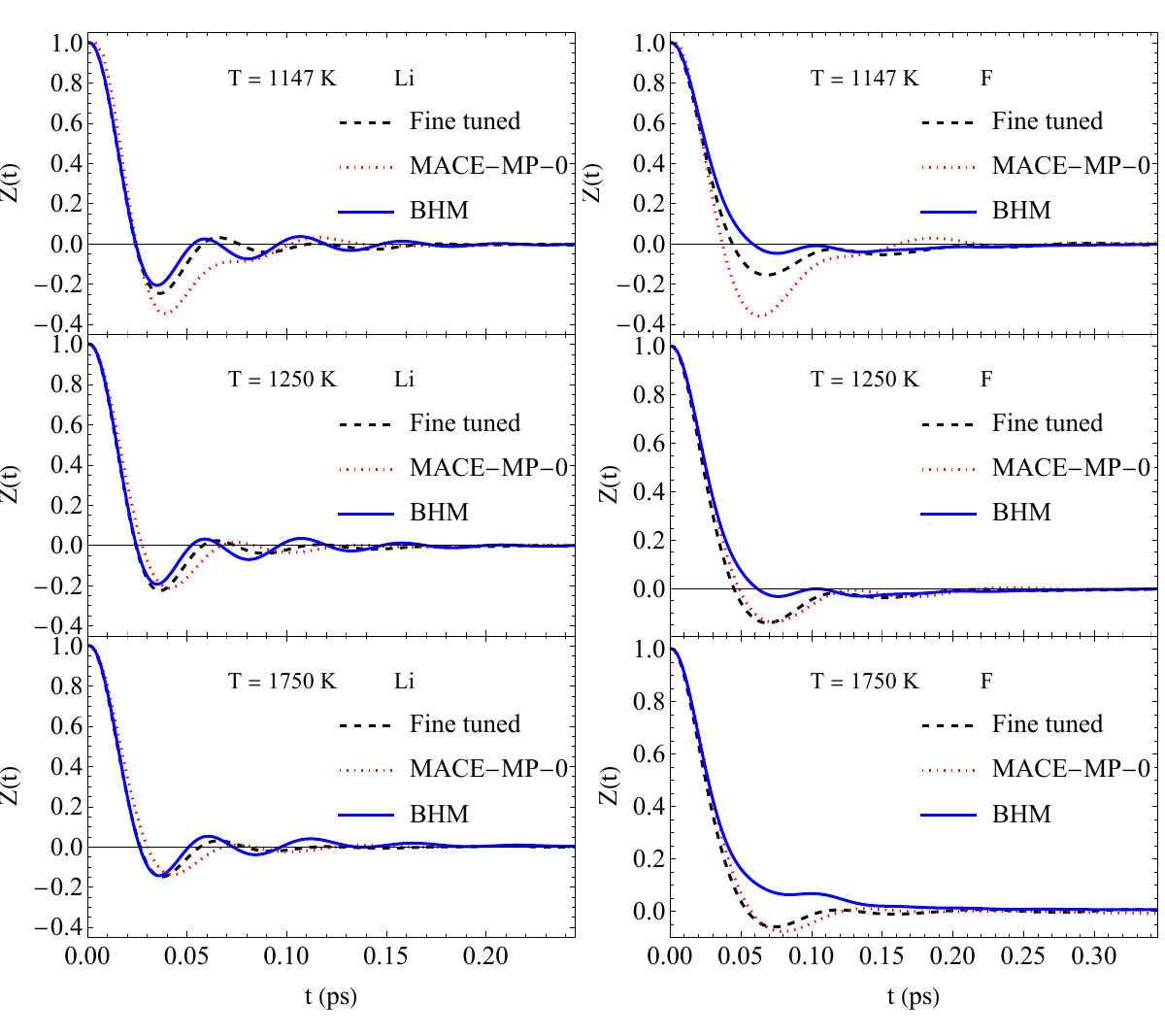}
    \caption{VAFs $Z(t)$ of Li atoms (left) and F atoms (right) for the BHM, MACE-MP-0b3, and fine-tuned potentials as a function of temperature.}
    \label{fig:vaf}
\end{figure*}

\begin{figure*}
    \centering
    \includegraphics[width=0.9\linewidth]{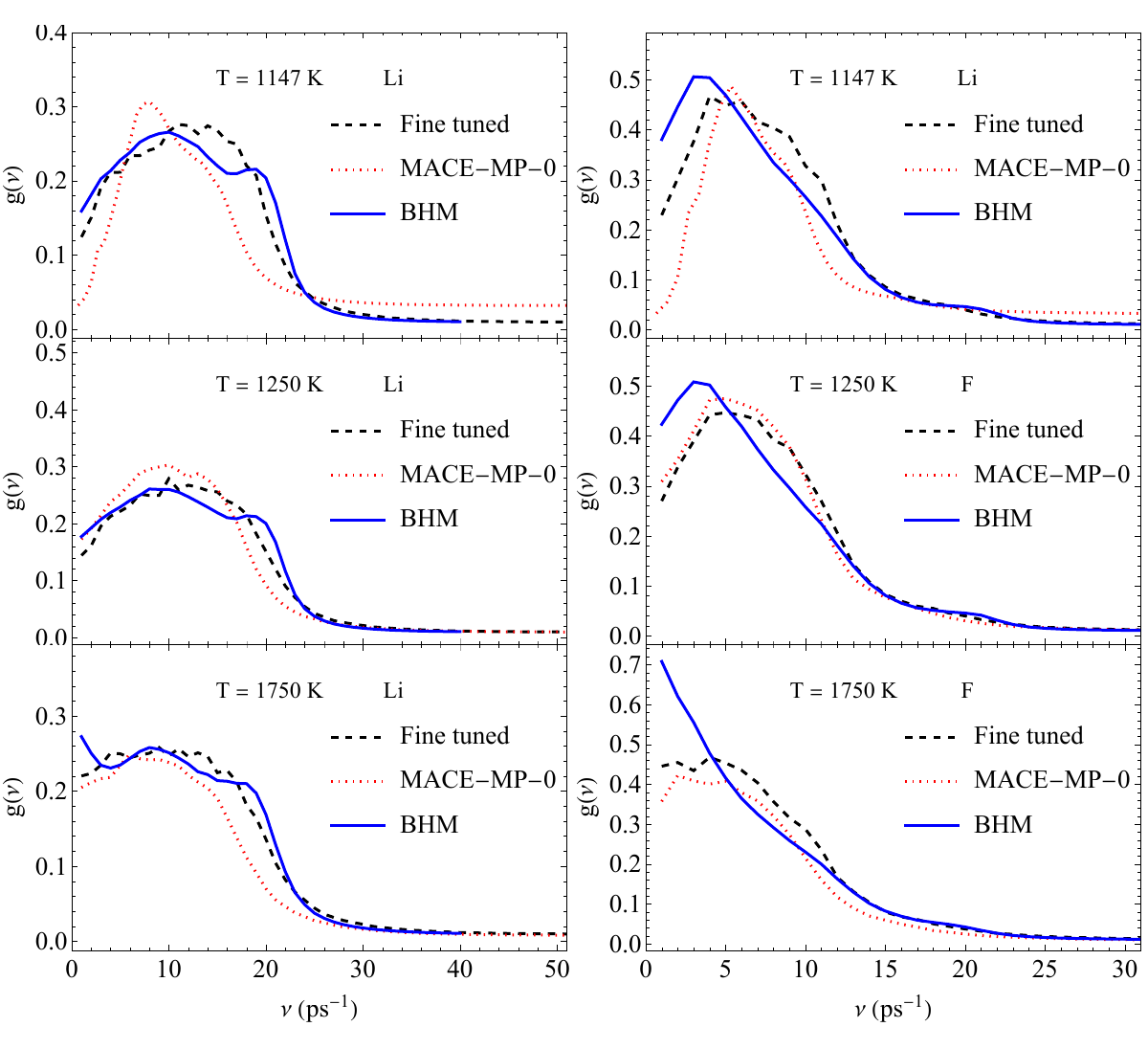}
    \caption{Densities of states $g(\nu)$ of Li atoms (left) and F atoms (right), calculated from the Fourier transform of the VAFs, for the BHM, MACE-MP-0b3, and fine-tuned potentials as a function of temperature.}
    \label{fig:dos}
\end{figure*}

Given the \textcolor{black}{improved} accuracy of the viscosity results of the MACE model we also examine the underlying microscopic structure and dynamics, \textcolor{black}{to compare the fundamentals of each model}. First we show the velocity autocorrelation functions (VAFs) $Z(t)$ of the BHM, MACE-MP-0b3, and fine-tuned models in figure \ref{fig:vaf}, and the resultant densities of states (DOS) $g(\nu)$ of each, calculated from the Fourier transform of $Z(t)$, in Fig. \ref{fig:dos}. We calculate these values just above the melting point \textcolor{black}{of the MACE-MP-0 LiF model, noting that this is over 50\% in excess of the absolute temperature of melting of the BHM model}. 
The results for the MACE potential are significantly different from those of the BHM model. \textcolor{black}{At 1147 K, the principal component of the motion of Li in the MACE-MP-0b3 model is somewhat reduced in frequency compared to the BHM model. Furthermore, the minima, particularly the first minima, of each ionic VAF is much deeper in the MACE-MP-0b3 model - so while the ionic oscillation decreases in frequency, it increases in amplitude and correlation time. Interestingly, the shapes of the anion and cation VACFs are qualitatively similar in the MACE model (albeit with the Li operating on a faster timescale), which is not at all the case in the BHM model. This implies that the MACE foundation model reduces the distinction between the ionic species. The deeper minima of the MACE model can be interpreted as a more pronounced solid-like component to the motion, with diffusive transits occurring less often. In other words, the Frenkel relaxation time \cite{Frenkel1955} $\tau$ will be larger than in the BHM model. At higher temperatures the minima depths in the Li VACFs are effectively the same in all three models. The solid-like component of the cation motion is therefore approximately equivalent, in this sense, between the classical potential and MLIP. The anion motion, meanwhile, decays Relating $\tau$ to the Maxwell relaxation time $ \tau_{\mathrm{M}} = \eta/G_{\infty}$, with $G_{\infty}$ the high frequency shear modulus, the increase in viscosity could be directly motivated with the increased solid-like component of (anion) motion. We note that the proximity of the MACE-MP-0b3 model to its melting point at 1147 K will be related to its more pronounced oscillatory motion.} 


\textcolor{black}{That said, the fine-tuned model, also featured in Fig. \ref{fig:vaf} somewhat everses these changes. Visually, the fine-tuned VAFs sit mostly ``between" the classical and foundation model VAFs, with the principal frequency exceeding that of the foundation model, but being exceeded by the classical model. Insofar as fine tuning drives a model towards the dynamics of an \textit{ab initio} MD simulation, we may say that fixed-charge classical models are too stiff, and that the foundation model causes excessive mode softening due to a lack of differentiation between anion and cation dynamics.}


\begin{figure}
    \centering
    \includegraphics[width=0.85\linewidth]{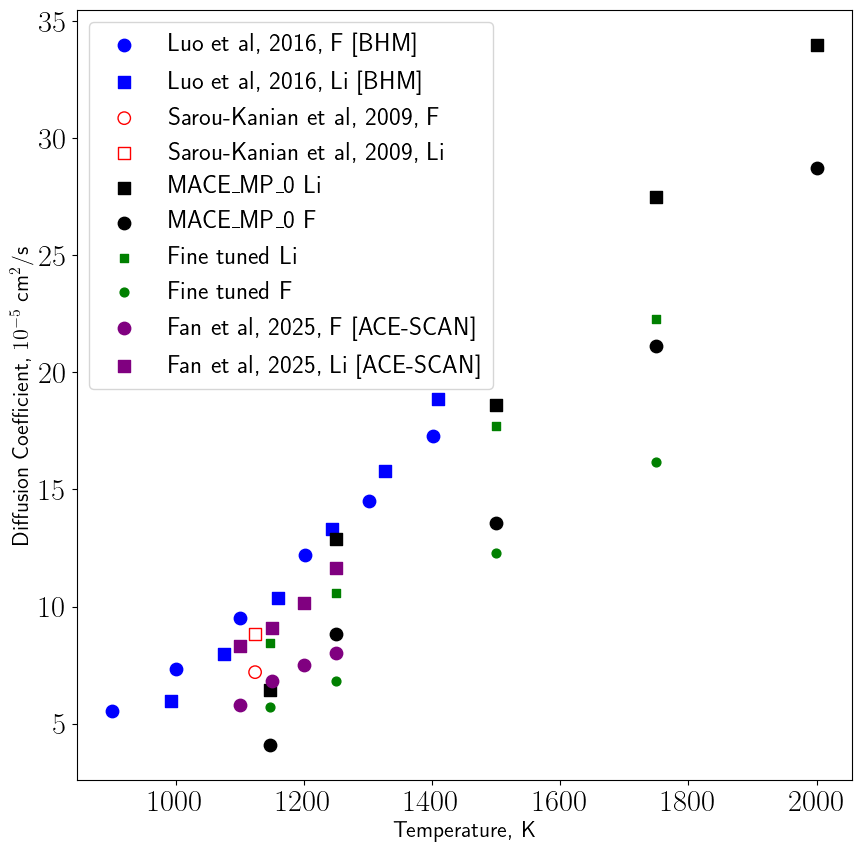}
    \caption{Diffusion coefficients calculated from the mean square displacement for MACE-MP-0b3 and the fine-tuned model compared to the BHM results of \cite{LUO2016203}, the experimental data taken from \cite{sarou2009diffusion}, and finally the simulation results of \cite{fan2025multistage}.}
    \label{fig:dc}
\end{figure}

\begin{figure*}
    \centering
    \includegraphics[width=\linewidth]{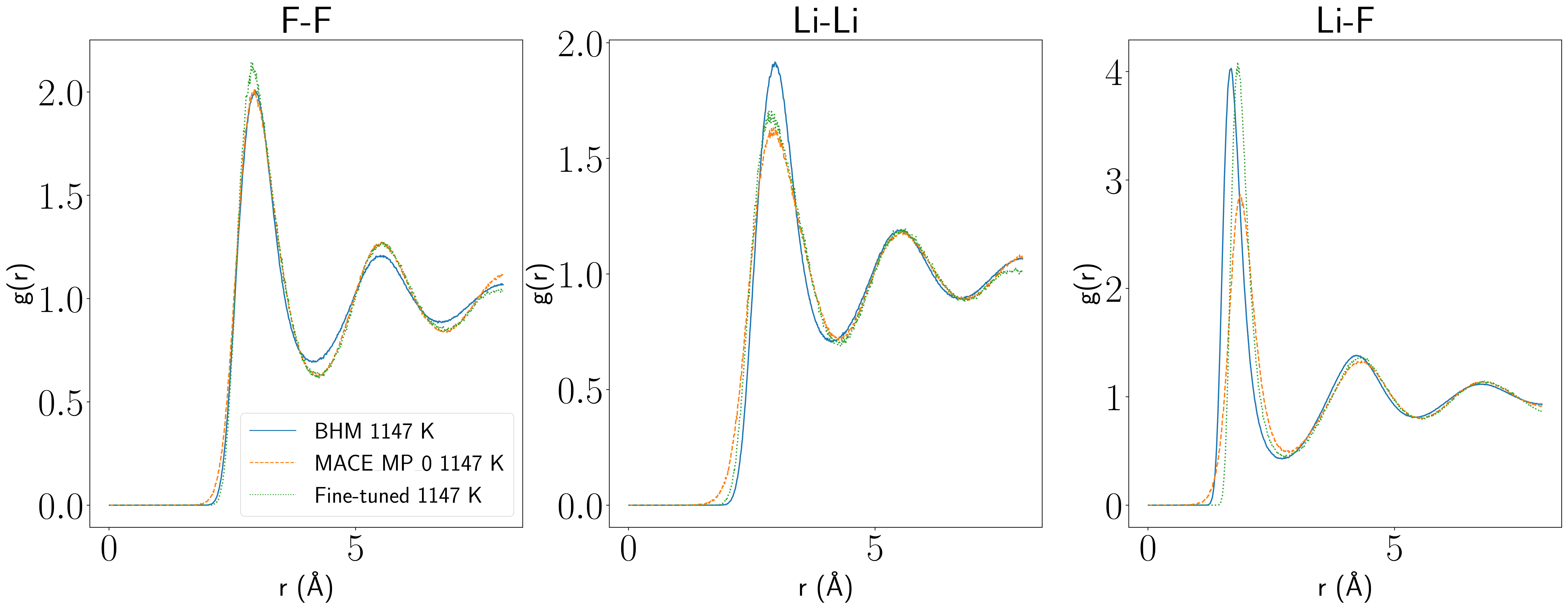}
    \caption{The radial distribution functions (RDFs) for the BHM, MACE\_MP\_0 and fine-tuned models. In each case we calculate the RDFs at 1147 K.}
    \label{fig:rdf}
\end{figure*}
We then plot the diffusion coefficients extracted from mean square displacements in Figure \ref{fig:dc}. We show our data alongside the experimental results for LiF reported by Sarou-Kanian \textit{et al.} \cite{sarou2009diffusion} the BHM results of Luo \textit{et al.} \cite{LUO2016203} and the ACE-SCAN data \cite{fan2025multistage}. The diffusion coefficient for the base and fine tuned models are generally in agreement with the available experimental data. Although with only one experimental scatter point conclusions are difficult to make. The fine-tuned model is a significant improvement and closer in agreement with ACE-SCAN as should be expected. 

Next we look at the partial RDFs $g(r)$ in Figure \ref{fig:rdf}
We show $g(r)$ for each pair \textcolor{black}{at 1147 K, the melting point of the MACE-MP-0b3 potential. Within the 7.5 \AA cutoff dictated by the small size of the MLIP systems, the first two peaks of each partial RDF are clearly visible, consistent with the expected results for LiF simulated with MD \cite{Ribeiro2003Chemla, abramo2020structure}. Peak positions in each RDF are broad agreement across the three models. The Li partial RDFs of the classical BHM model are considerably more pronounced than the MACE-MP-0 model - the peaks are both narrower and higher. The opposite is true for the F-F partial RDF, with the higher order coordination shells being broader in the classical model. In each RDF, the fine tuned model exhibits more pronounced peaks than the foundation model. A general comparison of the models is therefore as follows: the classical structures are fairly sharp, and the MACE-MP-0 foundation model broadens these coordination spheres. The fine-tuned MLIP then re-sharpens them, albeit with peaks occurring at somewhat different radial distances. We note the similar qualitative trends seen in analysis of the VACFs, with a sharp coordination environment being conducive to stiff vibrational modes.} 



\textcolor{black}{Nevertheless, the RDFs and VACFs of the BHM potential and fine-tuned MLIP are fairly similar. This stands in contrast to the rather pronounced difference in the viscosity of these models. This underscores the point that local equilibrium structures and single-particle motion do not fully determine the collective relaxation that gives rise to macroscopic viscosity. }

\begin{figure}
    \centering
    \includegraphics[width=0.999\linewidth]{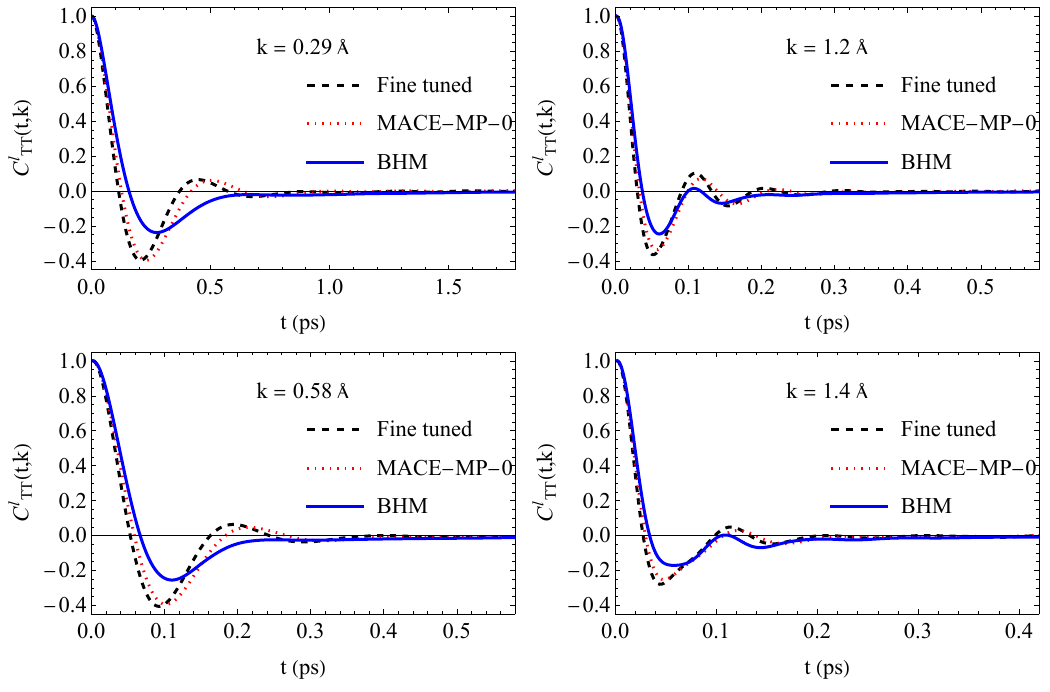}
    \caption{Normalised longitudinal total mass current autocorrelation functions $C^{\mathrm{l}}_{\mathrm{TT}}(t,k)$ in each model at four different wavenumbers.}
    \label{fig:ClongTTtk}
\end{figure}

\textcolor{black}{We conclude with a discussion of the collective modes operating in each model \textit{via} analysis of the current autocorrelation functions. We begin with the longitudinal total mass current autocorrelation function: $C^{\mathrm{l}}_{\mathrm{TT}}(t,k)$, which describes the decay of density waves as a function of wavelength - we plot this function (normalised) of each model at 1147 K in Fig. \ref{fig:ClongTTtk} at four different wavenumbers. Here we note that at longer wavelengths, the BHM model autocorrelates with a lower frequency than the MLIP model, while at shorter wavelengths the models become more similar. At all wavelengths, however, the total mass current decorrelates more quickly (the modes are more damped) in the classical model. In contrast, then, to single particle motion, the collective motion so far operates at higher frequencies in the MLIP models, and is less attenuated. The fine-tuned and foundation models are very similar at all wavelengths, except the shortest wavelengths where the decorrelation of the fine-tuned model is somewhat faster.}

\begin{figure}
    \centering
    \includegraphics[width=0.999\linewidth]{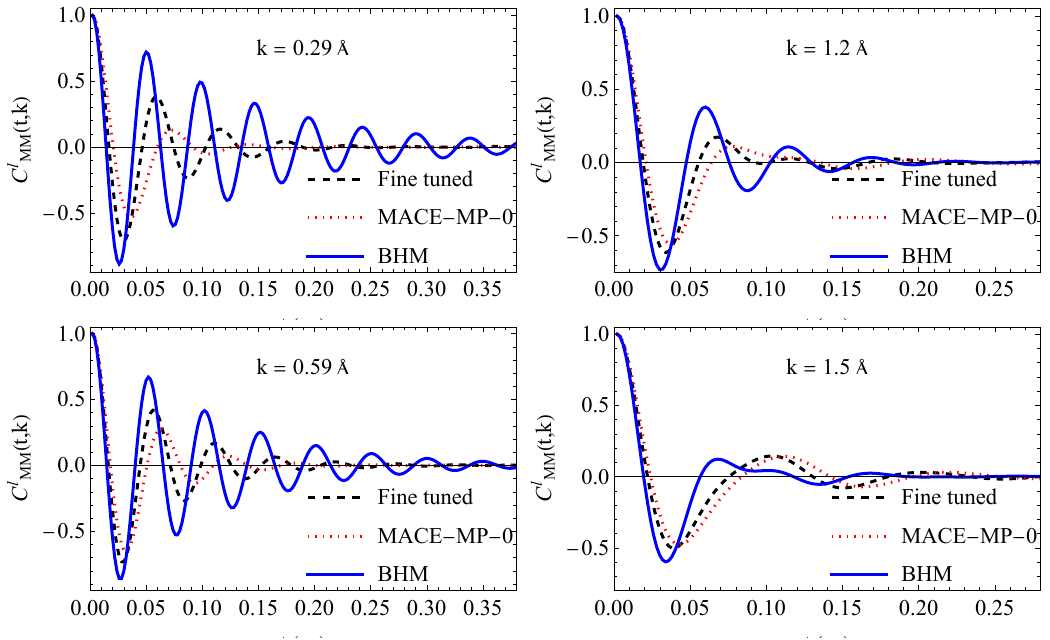}
    \caption{Normalised longitudinal mutual mass current autocorrelation functions $C^{\mathrm{l}}_{\mathrm{MM}}(t,k)$ in each model at four different wavenumbers.}
    \label{fig:ClongMMtk}
\end{figure}

\textcolor{black}{The longitudinal mutual mass current autocorrelation function, $C^{\mathrm{l}}_{\mathrm{MM}}(t,k)$, unlike $C^{\mathrm{l}}_{\mathrm{TT}}(t,k)$, does not describe propagating density modes at hydrodynamic wavelengths, but rather standing modes in partial densities. We plot these normalised functions of each model at 1147 K for the same four wavenumbers in Fig. \ref{fig:ClongMMtk}. Here, the model differences are much more pronounced, with the classical model operating at a higher frequency, and decaying significantly more slowly, at longer wavelengths. In each model, the autocorrelation function becomes less dominated by a single frequency and attenuates faster at shorter wavelengths, and indeed the classical and MLIP models become more similar to one another. It is in the mutual collective motion, or optic-like phonons, that the dynamical differences between the models are most pronounced. As the optic modes are related to charge fluctuations, we may relate the stiffness of fixed charge systems to these higher frequencies. As is typical, the long wavelength modes are higher in frequency, whereas dynamical disorder softens and increases attenuation of short wavelength modes, where mutual mass currents involve the relative motion of nearest neighbours. Here, as in the VACFs, the fine tuned model has behaviour (frequency and amplitude) between the foundation model and the classical model, with the latter being too stiff and under-attenuated, and the former being too soft and diffuse.}

\begin{figure}
    \centering
    \includegraphics[width=0.999\linewidth]{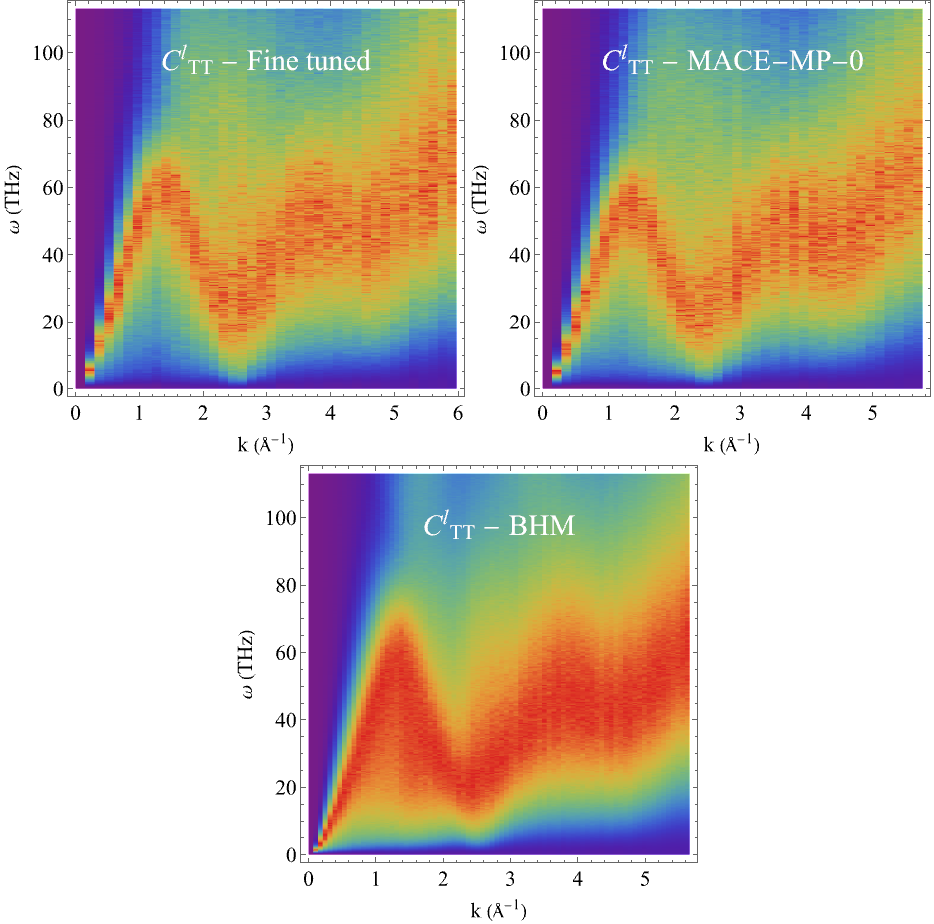}
    \caption{Normalised spectra of the total longitudinal mass current, $\tilde{C}_{\mathrm{TT}}^{\mathrm{l}}(\omega,k)$, of each model.}
    \label{fig:spectrum_tl}
\end{figure}

\textcolor{black}{We close with an examination of the normalised spectra, $\tilde{C}(\omega, k)$ of each current autocorrelation function, as this is more informative in order to discern the properties of the collective modes operating in each model. We begin with the total longitudinal spectrum $\tilde{C}_{\mathrm{TT}}^{\mathrm{l}}(\omega,k)$ in Fig. \ref{fig:spectrum_tl}, which describes the frequency-wavenumber relation of density waves within the liquid. In each model, at larger wavelengths, the spectra exhibit a clear linear relation, whose slope is the sound speed \cite{Balucani1994}, though we note that due to the small cell size, this speed will likely differ from that of the experimental hydrodynamic speed of sound. We note that the peaks, particularly at larger $k$, are very broad, despite superficial resemblance to crystalline dispersion curves. Both MLIP models exhibit very similar longitudinal total spectra, somewhat higher in frequency than the BHM model. The maxima and minima of the spectra of all models are in similar places, reinforcing our earlier observation that the classical and MLIP models have very similar longitudinal collective momentum modes except in the hydrodynamic limit.}

\begin{figure}
    \centering
    \includegraphics[width=0.999\linewidth]{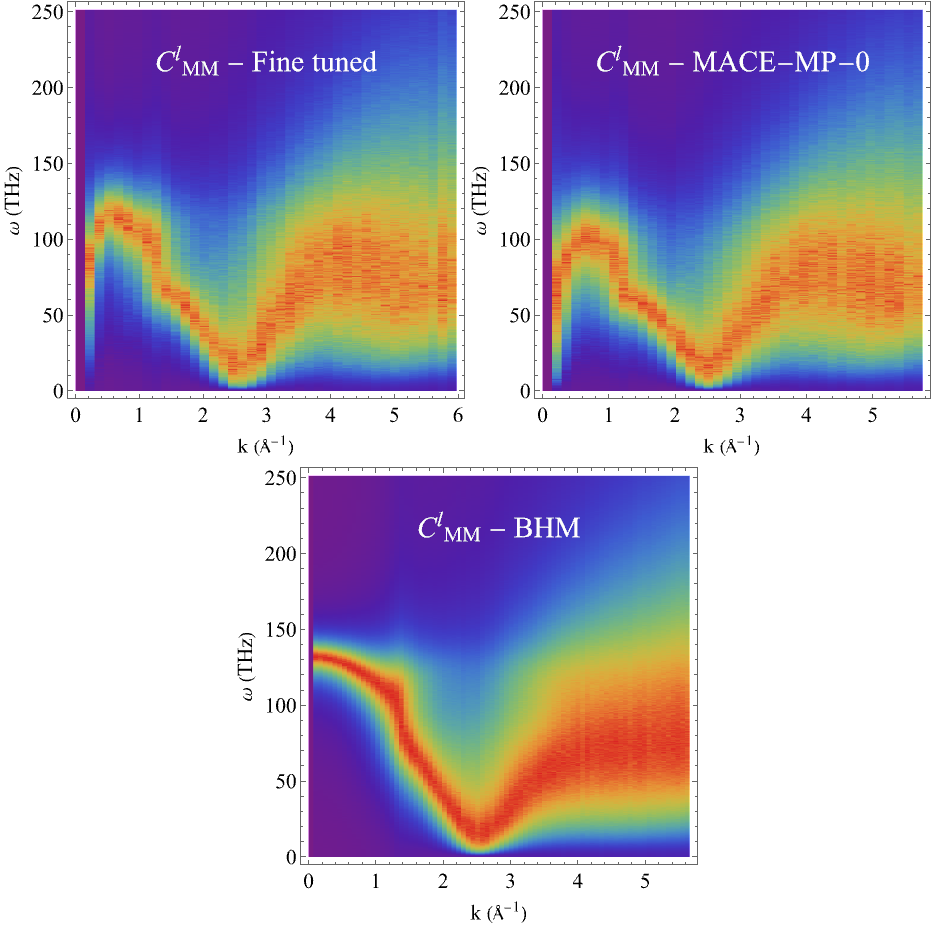}
    \caption{Normalised spectra of the mutual longitudinal mass current, $\tilde{C}_{\mathrm{MM}}^{\mathrm{l}}(\omega,k)$, of each model.}
    \label{fig:spectrum_ml}
\end{figure}

\textcolor{black}{In Fig. \ref{fig:spectrum_ml} we plot the spectra $\tilde{C}_{\mathrm{MM}}^l(\omega, k)$. While again the spectra are of a similar shape, the much higher frequency of the classical spectra, especially at low $k$, is clearly visible. This directly demonstrates the increased stiffness of charge current modes in the classical model, with the fine-tuned model lying between BHM and MACE-MP-0. Furthermore, the mutual spectra peaks are somewhat narrower in the BHM model than the MLIP models, while the reverse is true for the total spectra. The MACE models present an abnormally low frequency peak at the lowest wavelength (behaviour absent in the classical model) - this implies that the collective mutual anion-cation motion becomes higher in frequency at shorter wavelengths. The lack of long-range interactions / explicit inclusion of electrostatic charge in the MACE models seems a likely contributing factor to this behaviour. The reach of electrostatic forces beyond a few neighbour shells may allow the classical model to experience stiffer mutual mass modes. A full exploration of the effect of charge in MLIP models of this and similar systems is beyond the scope of this work.}

\begin{figure}
    \centering
    \includegraphics[width=0.999\linewidth]{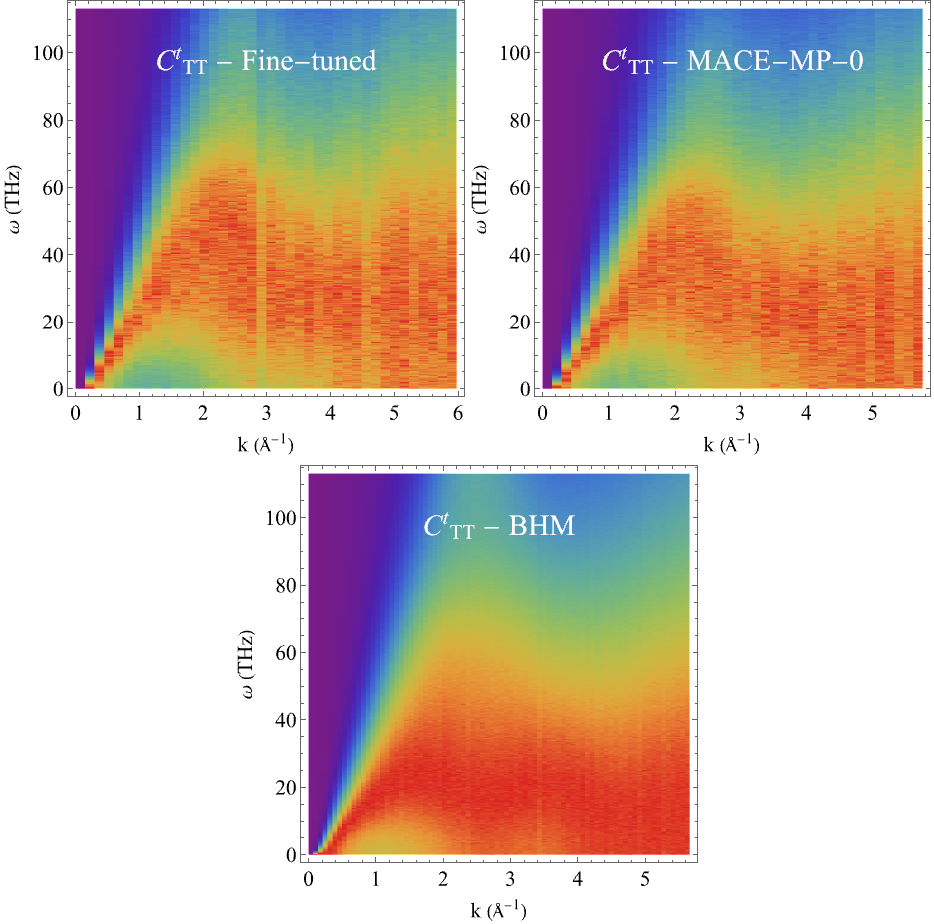}
    \caption{Normalised spectra of the total transverse mass current, $\tilde{C}_{\mathrm{TT}}^{\mathrm{t}}(\omega,k)$, of each model.}
    \label{fig:spectrum_tt}
\end{figure}

\begin{figure}
    \centering
    \includegraphics[width=0.999\linewidth]{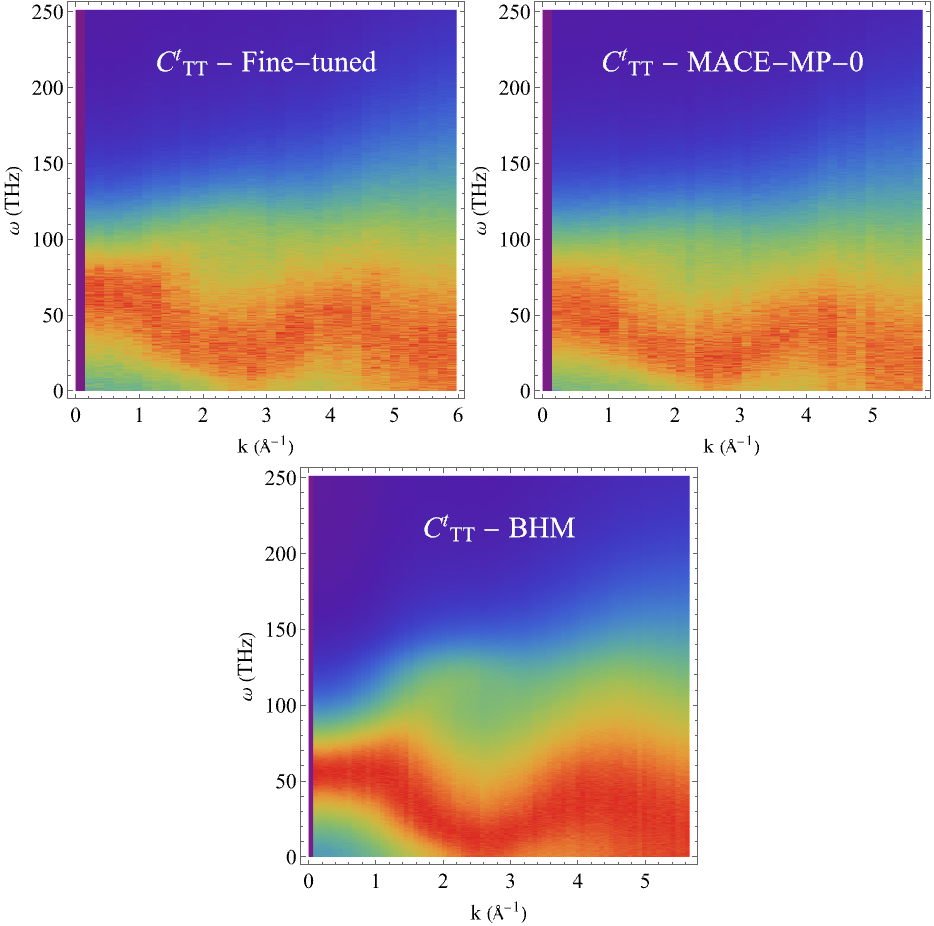}
    \caption{Normalised spectra of the mutual transverse mass current, $\tilde{C}_{\mathrm{MM}}^{\mathrm{t}}(\omega,k)$, of each model.}
    \label{fig:spectrum_mt}
\end{figure}

\textcolor{black}{Finally, we show the transverse total and mutual spectra $\tilde{C}_{\mathrm{TT}}^{\mathrm{t}}(\omega,k)$ and $\tilde{C}_{\mathrm{MM}}^{\mathrm{t}}(\omega,k)$ in Figs. \ref{fig:spectrum_tt} and \ref{fig:spectrum_mt}. Unlike in solids, transverse momentum modes do not propagate in the hydrodynamic in liquids \cite{Hansen2006}, leading to the existence of a ``gapped" wavenumber $k_{\mathrm{g}}$, related to the liquid relaxation time $\tau$ \cite{Frenkel1955}. In single component liquids, this causes the linear hydrodynamic regime of the dispersion curves to terminate at a $k_\mathrm{g}$ that is considerably above 0, unlike longitudinal dispersion curves that always terminate at $k = 0$ and $\omega = 0$. The gapped wavenumber differs considerably between the classical and MLIP models - $k_\mathrm{g} \approx 0.15$ \AA \ in the latter, $\approx 0.3$ \AA \ in the former. As mentioned above, the liquid relaxation time $\tau$ is often related to the Maxwell time $\tau_{\mathrm{M}} = \eta/G_\infty$ \cite{ropp}, and the decrease of $k_{\mathrm{g}}$ follows directly from the increased viscosity of the MLIP models. As expected, the transverse spectra are considerably broader than the longitudinal spectra, as atomic diffusion disrupts transverse modes, and the classical transverse branch is considerably lower in frequency than the MLIP branches. The mutual transverse mass spectra are more similar between the models, though the MLIP models have broader peaks - and the opposite is true for the total transverse mass currents - as was observed in the longitudinal spectra. The spectra peaks decrease with increasing $k$ from $k = 0$ in all models - the anomaly in $\tilde{C}_{\mathrm{MM}}^{\mathrm{l}}(\omega,k)$ is not present in $\tilde{C}_{\mathrm{MM}}^{\mathrm{t}}(\omega,k)$. This is curious, but as mentioned above, investigation is beyond the scope of this work.}

\textcolor{black}{Overall, the fixed charge BHM model results in more diffusive and  higher frequency single-ion motion, and  narrower, higher frequency mutual mass collective modes. Meanwhile, the total mass / momentum density modes are higher frequency in the MLIP models, and the increased viscosity is manifested in narrower transverse modes with a smaller $k_{\mathrm{g}}$. Though the transport, structure, and single-ion properties of the foundation and fine-tuned models qualitatively differ, the collective modes of each are nearly identical, though we note that the highly attenuated liquid collective modes results in signifiant peak broadening in the spectra, impeding differentiation.}

\textcolor{black}{These results elucidate the effect of rigid ion models the transport of liquids. While fixed formal charges result in fast, stiff dynamics of individual particles and their mutual, optic-like collective motion, they inhibit the sort of coherent, elastic-like motion that governs the transport of momentum, and therefore the viscosity. The short range many-body forces of MLIPs enhance this local elasticity, resulting in differences most pronounced at longer wavelengths. The effect of these short range forces, and the absence of long-range electrostatic forces, on long-wavelength dynamics is motivated by the curious ``optical" branches of the MACE collective modes.}

\section{Conclusions}
We have investigated the MD performance of \textcolor{black}{MACE MLIPs and classical interatomic potentials} of the molten salt LiF. \textcolor{black}{Out of the box, the MACE foundation model} is able to more accurately reproduce experimental properties of molten LiF, but at the cost of an increased computational load versus classical two-body potentials. \textcolor{black}{Model fine-tuning results in further changes to the viscosity, though given the spread in experimental data, it is hard to comment on whether this has resulted in improvement.}

Our main interest was in the prediction of shear viscosity\textcolor{black}{, and the dynamics governing it,} using the same methods as has been used in extant MD potentials (Buckingham and BHM) that are known to produce the correct trend but offset along the temperature axis for LiF. We have found that the MACE potential agrees well with experimental viscosities over a range of temperatures above melting, similarly to PIM results despite polarisation not being included. Both PIM and the MLIP results presented are parameterised by or trained on DFT data. The former improves due to its explicit treatment of electrostatic interactions, and for the MLIPs improvements come via their many-body nature.

\textcolor{black}{Amongst the Fumi and Tosi potentials, LiF is notable for its very poor prediction of the melting temperature. Using na\"ive methods - uniform heating rather than interface-based non-equilibrium MD simulations - results in a significant improvement to the melting line estimation. Given that a small system size and non-local heating method is only likely to overestimate the melting temperature, the improvement presented by the MLIPs are likely to persist the models are subjected to a more rigorous method of melting point estimation.} The large discrepancy between the melting points of classical potentials and experiments has previously been the given explanation for the offset viscosity values. \textcolor{black}{The MLIP models do address the simultaneous issues of underestimated melting temperatures and viscosities, with increased viscosity being correlated with increased melting temperature. However, the crude method employed will obfuscate the precise relationship between these two properties.}

Looking to the system's structure and dynamics, our results indicate that the MACE potential generates a broader coordination environment around each ion, reducing the stiffness of the aggregate interactions and thereby causing the decreased frequency in the oscillation of the Li atoms. The dynamics of the anions and cations are more similar in the MACE simulations, while in classical MD simulations they are quite distinct. \textcolor{black}{The fine-tuned model, generally speaking, exhibits behaviour between these extremes. The ``optic-like" collective modes in the MACE models are very different from the classical models, and exhibit a positive dispersion at low wavenumbers. This may be related to the implicit handling of charge in MACE models, and warrants future work.}

The far superior prediction of the viscosity and melting temperature of LiF indicate that the dynamics resulting the MACE potential, which differ qualitatively from those of classical potentials, are indispensable to the accurate simulation of this system. This increase in accuracy is however limited by the available experimental data. Our example is the extent of viscosity measurements across the temperature scales of the molten state. The viscosities as calculated by the PIM, MACE-MP-0b3, and the fine-tuned potential all sit within the range of experimental data available. The origin of the experimental range itself is not immediately clear. One possibility is the impact of impurities. This has been investigated recently for carbonates in NaCl-KCl \cite{liu2026effect} where a range of carbonate content results in a similar spread of viscosities. Another would be the details of the experiments themselves. The data we have found span a range of temporal and geographical contexts, with different experimental setups. Dedicated potentials for the simulation of other molten salts and their mixtures will help guide experimentation to address these gaps, and inform their application to industry.

\ack{This research utilised Queen Mary's Apocrita HPC facility, supported by QMUL Research-IT http://doi.org/10.5281/zenodo.438045, STFC Scientific Computing Department’s SCARF cluster and the Sulis Tier 2 HPC platform hosted by the Scientific Computing Research Technology Platform at the University of Warwick and funded by EPSRC GrantS EP/T022108/1 and the HPC Midlands+ consortium. Computing resources provided by STFC Scientific Computing Department’s SCARF cluster.}

\funding{We are grateful to EPSRC for grants EP/X011607/1 (H.L.D., M.W., C.C., K.T.), and EP/W029006/1 (H.L.D., K.T., A.M.E.). A.M.E. was also supported by the Ada Lovelace Centre at the Science and Technology Facilities Council (https://adalovelacecentre.ac.uk/), the Physical Sciences Data Infrastructure (https://psdi.ac.uk; jointly STFC and the University of Southampton) under grants EP/X032663/1 and EP/X032701/1, and EPSRC under grants EP/W026775/1 and EP/V028537/1.}

\roles{\textbf{H. L. Devereux}: Conceptualization; Data curation; Formal analysis; Investigation; Methodology; Software; Validation; Visualization; Writing - original draft; Writing - review \& editing, \textbf{M. Withington}: Conceptualization; Data curation; Formal analysis; Investigation; Methodology; Validation; Visualization; Writing - original draft; Writing - review \& editing, \textbf{C. Cockrell}: Conceptualization; Data curation; Formal analysis; Investigation; Methodology; Validation; Visualization; Writing - original draft; Writing - review \& editing, \textbf{K. Trachenko}: Conceptualization; Formal analysis; Investigation; Methodology; Validation; Writing - original draft; Writing - review \& editing; Funding acquisition; Supervision, \textbf{A. M. Elena}: Conceptualization; Data curation; Formal analysis; Investigation; Methodology; Software; Validation; Writing - original draft; Writing - review \& editing; Funding acquisition; Project administration; Supervision; Resources}

\data{The simulation code is available here \url{https://github.com/stfc/janus-core}, MACE-MP-0b3 and the MACE suite are available here \url{https://github.com/ACEsuit/mace/releases}, \url{https://github.com/ACEsuit/mace-foundations/releases}. Our fine-tuned model and training input data can be found here \url{https://data-collections.psdi.ac.uk/records/vtmb1-sr573}}

\bibliographystyle{unsrt}
\bibliography{bib.bib}
\end{document}